\documentclass[final,5p,times,twocolumn,preprint,compress]{elsarticle}
\usepackage{slashed}
\usepackage{eucal} 
\usepackage{bm}
\usepackage{amsmath}
\usepackage{amsthm}
\usepackage{amssymb}
\usepackage{graphicx}
\usepackage{amsfonts}
\usepackage{amscd}
\usepackage{amsxtra}
\usepackage{epstopdf}
\usepackage{dcolumn}
\usepackage{multirow}
\usepackage{physics}
\usepackage{subcaption}
\usepackage{dsfont}
\usepackage{cancel}
\usepackage{latexsym}
\usepackage[hypertexnames=false, colorlinks=true, citecolor=blue, urlcolor=blue, linkcolor=black]{hyperref}

\allowdisplaybreaks

\usepackage[normalem]{ulem}
\usepackage[dvipsnames]{xcolor}
\usepackage{array}
\usepackage{slashed}
\renewcommand{\sout}{\bgroup \color{red} \ULdepth=-.5ex \ULset}

\makeatletter

\begin{document}
\title{Mapping the transverse spin sum rule in position space} 
\begin{frontmatter}
\author[i]{C\'edric Lorc\'e}
\ead{cedric.lorce@polytechnique.edu}

\author[j]{Asmita Mukherjee}
\ead{asmita@phy.iitb.ac.in}

\author[j]{Ravi Singh}
\ead{ravi.singh298@iitb.ac.in}

\author[i]{Ho-Yeon Won}
\ead{hoyeon.won@polytechnique.edu}

\address[i]{CPHT, CNRS, \'Ecole polytechnique, Institut Polytechnique de Paris, 91120 Palaiseau, France}

\address[j]{Department of Physics,
Indian Institute of Technology Bombay, Powai, Mumbai 400076, India}

\begin{abstract}
We discuss in detail the relativistic spatial distribution of transverse angular momentum, including both orbital and intrinsic spin contributions.
Using the quantum phase-space formalism, we begin with the definition of the three-dimensional spatial distributions of transverse orbital angular momentum and intrinsic spin in a generic Lorentz frame.
By integrating these three-dimensional spatial distributions over the longitudinal axis, we derive for the first time the relativistic spatial distributions of transverse orbital angular momentum, intrinsic spin, and total angular momentum for spin-0 and spin-1/2 targets in the transverse plane.
We verify the transverse spin sum rule about the relativistic center of spin for spin-0 and spin-1/2 systems, and find that the transverse total angular momentum distribution is non-trivial, even for spin-0 targets. 
We also show how the distributions of transverse orbital angular momentum, intrinsic spin, and total angular momentum change with the target momentum.

\end{abstract}

\end{frontmatter}

\section{Introduction}
One of the most fundamental questions in particle physics is to understand the origin of the nucleon's spin.
In general, studying the spin structure of the nucleon involves multiple aspects which can be broadly categorized into:
(i) determining the individual contributions of the constituents' orbital angular momentum (OAM) and intrinsic spin to the total angular momentum (TAM) and
(ii) analyzing the angular momentum (AM) distributions in both position and momentum space. 
Although significant progress has been made regarding the first aspect through numerous experimental measurements, see e.g.~\cite{EuropeanMuon:1987isl, EuropeanMuon:1989yki,  COMPASS:2005qpp, Boyle:2006ab, Kuhn:2008sy, deFlorian:2014yva}, the spatial distribution of AM remains relatively underexplored.
These questions are of utmost importance and constitute one of the pillars of the Electron-Ion Collider (EIC) project~\cite{Accardi:2012qut,Aschenauer:2017jsk,AbdulKhalek:2021gbh} in the USA.

The spin sum rule for longitudinal AM in a longitudinally polarized nucleon is well-established~\cite{Jaffe:1989jz,Ji:1996ek}, while transverse spin sum rules in a transversely polarized nucleon have been debated over the years~\cite{Harindranath:2001rc,Burkardt:2002hr,Bakker:2004ib,Burkardt:2005hp,Leader:2011cr,Ji:2012sj,Ji:2012vj,Leader:2012md,Harindranath:2012wn,Leader:2012ar,Hatta:2012jm,Harindranath:2013goa,Ji:2013tva,Ji:2020hii,Lorce:2021gxs}.
In essence, a significant challenge arises from the non-commutativity between the transverse AM operator and the longitudinal Lorentz boost operator~\cite{Lorce:2018zpf}.
This non-commutativity leads to frame-dependent effects, resulting in differing outcomes in the literature. 
In a recent paper~\cite{Ji:2020hii}, the authors derived a transverse spin sum rule in a transversely polarized nucleon and compared the expressions obtained from the expectation value of the transverse AM with the Pauli-Lubański pseudovector~\cite{Lubanski:1942min,Lubanski:1942log}. 
They interpreted the mismatched terms as the contribution from the center-of-mass motion of the moving nucleon, and removed these terms by hand without providing a clear physical justification.
This result also seemed to contradict the findings of Ref.~\cite{Leader:2011cr}.
The latter discrepancy has recently been resolved in Ref.~\cite{Lorce:2021gxs} through an investigation of the role played by different pivots in quantum field theory (QFT).

The spatial distributions of AM are derived from the Fourier transform of the matrix elements of the energy-momentum tensor (EMT), and are traditionally defined in two reference frames:
the Breit frame (BF)~\cite{Polyakov:2002yz} in instant form (IF) coordinates and the Drell-Yan frame (DYF) in light-front (LF) coordinates~\cite{Lorce:2018egm,Freese:2021czn}. 
The BF is normally used to define three-dimensional (3D) spatial distributions, but their interpretation as genuine densities is spoiled by relativistic recoil corrections.
Alternatively, the DYF provides access to spatial distributions in two-dimensional (2D) position space with a proper density interpretation, thanks to the Galilean symmetry of the LF formalism in the transverse plane~\cite{Burkardt:2002hr,Miller:2010nz}. 
This Galilean symmetry can, alternatively, be achieved in the infinite-momentum frame (IMF) in the IF formalism~\cite{Susskind:1967rg}. 
To interpolate between the BF and the IMF, the elastic frame (EF) was introduced in ~\cite{Lorce:2018egm,Lorce:2017wkb} within the quantum phase-space formalism~\cite{Wigner:1932eb, Hillery:1983ms}, where the relativistic spatial distributions are interpreted as quasi-densities in the Wigner sense.

Despite these efforts, the transverse OAM distribution was not studied in the EF due to a key incompatibility: it requires differentiation with respect to longitudinal momentum transfer, $\Delta^z$, while the EF enforces $\Delta^z=0$ by its definition. 
The transverse intrinsic spin distribution has only been studied in the EF for an unpolarized target~\cite{Lorce:2018egm}. 
To remedy the limitations in defining relativistic spatial distributions in the transverse plane, we consider a generic frame (GF) in the 3D space.
The GF provides an interpolation between the BF and the IMF in the 3D space. While the 3D GF requires non-zero target momentum and thus violates the elastic condition $\Delta^0=0$, integrating the 3D relativistic spatial distributions over the longitudinal axis both yields the relativistic spatial distributions of transverse OAM in the transverse plane and projects the 3D GF onto the 2D EF, thereby restoring the elastic condition.
We also define the relativistic spatial distribution of transverse intrinsic spin using the generalized spin operator~\cite{Lorce:2017wkb}.

This article presents, for the first time, a detailed study of the relativistic spatial distributions of transverse OAM and intrinsic spin for spin-0 targets, and for both unpolarized and transversely polarized spin-1/2 targets.
In the case of longitudinally polarized spin-1/2 targets, 
the transverse AM distributions coincide with those of unpolarized targets (see discussion in Sec.~\ref{sec.3}).
This paper is organized as follows.
In Sec.~\ref{sec.2}, we briefly review the generalized AM tensor and asymmetric EMT operators in QCD, and define their matrix element in terms of a suitable parametrization.
In Sec.~\ref{sec.3}, we introduce the GF to define the relativistic spatial distributions of transverse OAM and intrinsic spin in the 3D space 
and derive the 2D spatial distributions of spin-0 and spin-1/2 targets in the transverse plane by integrating over the longitudinal axis.  
Furthermore, we verify that the transverse spin sum rules of spin-0 and spin-1/2 targets are satisfied, and present numerical results illustrating the behavior of transverse OAM, intrinsic spin, and TAM distributions in spin-0 and spin-1/2 targets for various values of the target momentum.

\section{Generalized Angular Momentum and Energy-Momentum Tensor Operators \label{sec.2}}
\subsection{Angular Momentum in QCD}
In field theory, 
the generalized AM tensor operator $\hat{M}^{\mu\alpha\beta}$
is derived from the symmetry under the Lorentz transformation,
and is given by the sum of two contributions as
\begin{align}
    \hat{M}^{\mu \alpha \beta}
  = \hat{L}^{\mu \alpha \beta}
  + \hat{S}^{\mu \alpha \beta},
\label{generalizedAM}
\end{align}
where each contribution is antisymmetric under $\alpha\leftrightarrow\beta$.
In QCD, the first contribution is the generalized OAM tensor operator, expressed in terms of the conserved EMT, and the other is the generalized intrinsic spin operator~\cite{Leader:2013jra}
\begin{align}
    \hat{L}^{\mu \alpha \beta} 
& = x^{\alpha} 
    \hat{T}^{\mu \beta} 
  - x^\beta 
    \hat{T}^{\mu \alpha},\cr
    \hat{S}^{\mu\alpha\beta}
& = \frac{1}{2}
    \epsilon^{\mu\alpha\beta\lambda} \bar{\psi} \gamma_{\lambda} \gamma_{5} \psi,
\end{align}
with the convention $\epsilon_{0123}=1$.
The generalized AM tensor operators provide 
the TAM, OAM, and intrinsic spin operators, 
as follows: 
\begin{align}\label{AMcharges}
    \hat{J}^{i} 
& = \frac{1}{2}
    \epsilon^{ijk}
    \int d^3x \;
    \hat{M}^{0jk}, \cr
    \hat{L}^{i} 
& = \frac{1}{2}
    \epsilon^{ijk}
    \int d^3x \;
    \hat{L}^{0jk},
    \hspace{0.5cm}
    \hat{S}^{i} 
  = \frac{1}{2}
    \epsilon^{ijk}
    \int d^3x \;
    \hat{S}^{0jk}.
\end{align}

\subsection{Energy-Momentum Tensor in QCD}
In the present work, we adopt the local, gauge-invariant and asymmetric EMT operator~\cite{Leader:2013jra}. 
In QCD, 
it is given by the sum of gauge-invariant 
quark and gluon contributions as follows:
\begin{align}
    \hat{T}^{\mu\nu}
& = \hat{T}_{q}^{\mu\nu}
  + \hat{T}_{G}^{\mu\nu}
\end{align}
with
\begin{align}
    \hat{T}_{q}^{\mu\nu}
& = \bar{\psi} 
    \gamma^{\mu}    
    \frac{i}{2}  \overleftrightarrow{D}^{\nu} 
    \psi, \\
    \hat{T}_{G}^{\mu\nu}
& = 
  - 2 \mathrm{Tr}\;
    F^{\mu\lambda}  F_{\phantom{a}\lambda}^{\nu}
  + \frac{1}{2}
    g^{\mu\nu}\,
    \mathrm{Tr}\;
    F^{\lambda\rho}  F_{\lambda\rho}.
\label{EMTcurrent}
\end{align}
$\overleftrightarrow{D}^{\mu}=\overrightarrow{\partial}^{\mu}-\overleftarrow{\partial}^{\mu}-2igA^{\mu}$ and 
$F^{\mu\nu}=\partial^{\mu}A^{\nu}
-\partial^{\nu}A^{\mu}-ig[A^{\mu},A^{\nu}]$ denote the covariant derivative and gluon field-strength tensor, respectively.
The antisymmetric part of the quark EMT is related to the quark axial-vector current operator using the QCD equation of motion
as follows~\cite{Lorce:2017wkb,Leader:2013jra}:
\begin{align}
    \bar{\psi}
    \gamma^{[\alpha} 
    i\overleftrightarrow{D}^{\beta]}
    \psi
& = - 2 \partial_\mu S_{q}^{\mu \alpha \beta}
  = - \epsilon^{\alpha\beta\mu \nu}
    \partial_{\mu}
    \left(
    \bar{\psi}
    \gamma_{\nu}\gamma_{5}
    \psi
    \right),
\label{antiEMT}
\end{align}
where $a^{[\mu}b^{\nu]}=a^{\mu}b^{\nu}-a^{\nu}b^{\mu}$.

\subsection{Matrix Elements of the Energy-Momentum Tensor and Generalized Spin Operator}
We denote the matrix elements 
of some operator $\hat{O}(x)$ as follows
\begin{align}
    \mathcal{M}_{s^{\prime}s}
    [\hat{O}]
& = \mel{p^{\prime},s^{\prime}}{\hat{O}(0)}{p,s} .
\end{align}
The four-momentum states are normalized as 
$\braket{p^{\prime},s^{\prime}}{p,s}=2p^0
\left(2\pi\right)^{3}\delta^{(3)}\left(\bm{p}^{\prime}-\bm{p}\right)\delta_{s^{\prime}s}$, where 
$s \text{ and } s^{\prime}$ stand for the canonical spin polarizations of 
initial and final states.
For a spin-0 target, 
the matrix elements of the EMT are parametrized
in terms of three form factors~\cite{Pagels:1966zza,Hudson:2017xug}
\begin{align}
&   \mathcal{M}
    [\hat{T}_{a}^{\mu \nu}] 
  = 2P^{\mu}P^{\nu}
    A_{a} 
  + \frac{\Delta^{\mu}\Delta^{\nu}
  - g^{\mu\nu}\Delta^{2}}{2}  
    D_{a}
  + 2M^{2} g^{\mu\nu} 
    \bar{C}_{a},
\label{EMT_spin0}
\end{align}
where $a=q,G$ and $M$ is the target mass.
When the index $a$ is omitted on a form factor $F$, it means that we summed over the quark and gluon contributions, namely $F=F_{q}+F_{G}$. For convenience, we introduced the average momentum $P=(p'+p)/2$ and momentum transfer $\Delta=p'-p$ The mass shell condition for initial and final states, $p^{2}=p^{\prime2}=M^{2}$, leads to
\begin{align}
    P^{2}
  = M^{2}
  - \frac{\Delta^{2}}{4}
  = M^{2}
    \left(1+\tau\right)
    \quad 
    \textrm{and}
    \quad P\cdot\Delta=0
\end{align}
with the dimensionless Lorentz-invariant variable 
$\tau=-\frac{\Delta^{2}}{4M^{2}}$.

Similarly, 
the matrix elements of the EMT for a spin-1/2 target are parametrized 
in terms of ﬁve form factors~\cite{Bakker:2004ib,Leader:2013jra,Burkert:2023wzr}
\begin{align}
    \mathcal{M}_{s^{\prime}s}
    [\hat{T}_{a}^{\mu \nu}] 
& = \bar{u}'
    \Bigg[
    \frac{P^{\mu}P^{\nu}}{M}  
    A_{a}
  + \frac{\Delta^{\mu}\Delta^{\nu}-g^{\mu\nu}\Delta^{2}}{4M}  
    D_{a}   
  + M g^{\mu\nu} \bar{C}_{a}\cr
& + \frac{iP^{\{\mu}\sigma^{\nu\}\rho}\Delta_{\rho}}{2M}  
    J_{a}
  - \frac{iP^{[\mu}\sigma^{\nu]\rho}\Delta_{\rho}}{2M}  
    S_{a}
    \Bigg]
    u, 
\label{EMT}
\end{align}
where $a^{\{\mu}b^{\nu\}}=a^{\mu}b^{\nu}+a^{\nu}b^{\mu}$.
The Dirac spinors are represented as $u:=u(p,s)$ and $\bar{u}':=\bar{u}(p^{\prime},s^{\prime})$ with the normalization $\bar{u}(p,s)u(p,s)=2M$. The matrix elements of the quark generalized spin operator for a spin-1/2 target are parametrized as
\begin{align}
&   \mathcal{M}_{s^{\prime}s}
    [\hat{S}_{q}^{\mu\alpha\beta}] 
  = \frac{1}{2} \epsilon^{\mu\alpha\beta\lambda}
    \bar{u}' 
    \left[
    \gamma_{\lambda} \gamma_{5}
    G_{A}^{q}  
  + \frac{\Delta_{\lambda}\gamma_{5}}{2M}  
    G_{P}^{q}   
    \right]
    u.
\label{axial}
\end{align}
In Eqs.~\eqref{EMT_spin0}, \eqref{EMT} and \eqref{axial}, 
the form factors $F:=F(t)$ with $F=A,D,\bar{C},J,S,G_{A},G_{P}$
are real-valued functions of the momentum
transfer squared, $t=\Delta^{2}$. 
Using the relation in Eq.~\eqref{antiEMT}, 
one obtains the relation between the spin and axial-vector
form factors as~\cite{Lorce:2017wkb}
\begin{align}\label{spinrelation}
    S_{q}   
  = \frac{1}{2}
    G_{A}^{q},
    \hspace{1cm}
    S_{G}
  = 0.
\end{align}
The gluon contribution to the spin form factor is constrained to vanish, as there is no known way to decompose the gluon TAM into spin and OAM contributions in a local and gauge-invariant way~\cite{Leader:2013jra}.
As a result of the Poincar\'e symmetry,
the total EMT form factors are constrained as~\cite{Ji:1996ek,Leader:2013jra,Teryaev:1999su,Lowdon:2017idv,Cotogno:2019xcl}
\begin{align}\label{GFFconstraints}
    A(0)
& = 1, 
    \hspace{0.4cm}
    J(0)
  = \frac{1}{2},
    \hspace{0.4cm}
    \bar{C}(t)
  = 0.
\end{align}
In contrast, $D(0)$ is not fixed by any symmetry,
but is conjectured to be negative in QCD motivated by stability arguments~\cite{Burkert:2023wzr,Perevalova:2016dln,Polyakov:2018zvc,Lorce:2025oot}.

In the present work, we utilize the simple multipole Ansatz for the form factors
$A,D,J,S$ fitted to Lattice QCD calculations given in Ref.~\cite{Hackett:2023nkr} for the pion and in Ref.~\cite{Hackett:2023rif} for the nucleon,
to conduct a numerical analysis and illustrate our results.
For $G_{P}$, 
we take into account the pion-pole dominance and the dipole Ansatz 
from Refs.~\cite{Alexandrou:2021wzv,Chen:2024oxx}.

\section{Angular Momentum Distributions \label{sec.3}}

\subsection{Spatial Distribution in the Generic Frame}

We consider the GF in the 3D space to compute the matrix elements. 
The kinematic variables are given by 
\begin{align}
    P
& = \left(P^{0},\bm{0}_{\perp},P^{z}\right),
    \hspace{0.5cm}
    \Delta
  = \left(\Delta^{0},\bm{\Delta}_{\perp},\Delta^{z}\right),
\end{align} 
where we chose for convenience the direction of average momentum along the $z$-axis, i.e.,~$\bm{P}=P^{z}\bm{e}^{z}$. 
We have to pay the price 
for the finite average momentum in the 3D space, 
that is, we lose the elastic condition as 
\begin{align}
    \Delta^{0}
  = \frac{\bm{P}\cdot\bm{\Delta}}{P^{0}}
    \neq 0.
\end{align}

Following the quantum phase-space formalism~\cite{Lorce:2018egm,Won:2025dgc,Lorce:2020onh,Chen:2022smg,Chen:2023dxp}, 
we define the 3D spatial distributions in the GF, 
localized in the Wigner sense around 
the average position $R$ and average momentum $P$ as follows:
\begin{align}
    \expval{\hat{O}\left(\bm{r}\right)}_{\bm{R},\bm{P}}^{s^{\prime},s}
  = \int \frac{d^{3}\Delta}{\left(2\pi\right)^{3}}\;
    e^{-i\bm{\Delta}\cdot\left(\bm{r}-\bm{R}\right)}\,
    \langle\langle 
    \hat{O}
    \rangle\rangle,
\end{align}
where we used the following notation
\begin{align}
    \langle\langle 
    \hat{O}
    \rangle\rangle
& = \left.
    \frac{\mathcal{M}_{s^{\prime}s}
    [\hat{O}] }
    {2\sqrt{p^{\prime0}p^{0}}}\right|_{\mathrm{GF}}.
\end{align}
In this work, we have set $r^{0}=0$ and $ R^0=0$.

\subsection{Definition of Angular Momentum Distributions \label{sec.3.2}}
The internal OAM distribution in the 3D GF is given by
\begin{align}
    L_{a}^{i}
    \left(\bm{r},P^z;s^{\prime},s\right)
& = \epsilon^{ijk} 
    r^{j} 
    \expval{\hat{T}_{a}^{0k} \left(\bm{r}\right)}_{\bm{0},\bm{P}}^{s^{\prime},s}, 
    \cr
& = \epsilon^{ijk} 
    \int \frac{d^{3}\Delta}{(2\pi)^{3}}\, 
    e^{ -i\bm{\Delta}\cdot\bm{r}}
    \left[ 
  - i \frac{\partial \langle\langle 
    \hat{T}_{a}^{0k}
    \rangle\rangle}{\partial \Delta^{j}}  
    \right].
    \label{Lk2}
\end{align}
Note that we set $\bm{R}=\bm{0}$ to identify the coordinate $\bm{r}$ with the position relative to the ``canonical'' center of the system~\cite{Jaffe:1989jz,Bakker:2004ib,Leader:2013jra,Shore:1999be}.  The canonical center is also known as the center of spin, in the sense that the magnitude of the total internal AM of a spin-$j$ target calculated relative to that point is always $j$. We adopted this definition because it is the only one that corresponds to TAM operators that satisfy the $su(2)$ algebra for arbitrary values of the target momentum~\cite{Lorce:2021gxs,Lorce:2018zpf}. Alternative definitions based on other choices for the pivot will, however, be investigated in a future work~\cite{Lorce:2025ll}.

The intrinsic spin distribution is defined in the Wigner sense as
\begin{align}
    S_{a}^{i}
    \left(\bm{r},P^{z};s^{\prime},s\right)
&= \frac{1}{2}
    \epsilon^{ijk}
    \expval{\hat{S}_{a}^{0jk} \left(\bm{r}\right)}_{\bm{0},\bm{P}}^{s^{\prime},s},
    \label{Sk2} 
\end{align}
or more explicitly
\begin{equation}
\begin{aligned}
  S_{q}^{i}
    \left(\bm{r},P^{z};s^{\prime},s\right)&=\frac{1}{2} \expval{\bar\psi\left(\bm{r}\right)\gamma^i\psi\left(\bm{r}\right)}_{\bm{0},\bm{P}}^{s^{\prime},s},\\    S_{G}^{i}
    \left(\bm{r},P^{z};s^{\prime},s\right)&=0.
    \end{aligned}
\end{equation}

\subsection{Limitations of the Relativistic 3D Spatial Distribution}
As stressed in the introduction, the relativistic spatial distribution of transverse OAM cannot be directly derived in the EF. 
Thus, one must consider the transverse OAM distribution in the 3D position space.
However, our understanding of the status of 3D spatial distributions remains more limited compared to the 2D case, see e.g.~\cite{Burkardt:2002hr,Lorce:2018egm,Freese:2021czn,Won:2025dgc,Lorce:2020onh,Chen:2022smg,Chen:2023dxp,Miller:2009sg,Freese:2021mzg,Panteleeva:2022khw,Epelbaum:2022fjc,Carlson:2022eps,Alharazin:2022xvp,Panteleeva:2023evj}.

Interpreting spatial distributions as physical densities requires the hadronic wave packet to be well localized~\cite{Miller:2009sg,Yennie:1957skg,Kelly:2002if,Burkardt:2000za,Belitsky:2003nz,Jaffe:2020ebz}.
Within the quantum phase-space formalism, relativistic spatial distributions are obtained in the EF and turn into probabilistic densities in the IMF, as a result of Galilean symmetry in the transverse plane~\cite{Chen:2023dxp}. 
In contrast, in the 3D GF, the validity of such localization has not yet been firmly established within the phase-space formalism.

Due to such limitations, in the present work we concentrate on the 2D spatial distributions obtained by integrating the 3D ones derived in the GF over the longitudinal direction.
A detailed analysis of 3D GF distributions is left for future work.

\subsection{Spin-0 Target \label{sec.3.3}}
The generalized AM tensor operator in Eq.~\eqref{generalizedAM} and its decomposition into orbital and intrinsic spin contributions applies in QCD to targets of arbitrary spin~\cite{Leader:2013jra}. 
However,  for a spin-0 target, the TAM is entirely determined by the OAM, as the intrinsic spin vanishes in this case (there is no axial-vector current for a spin-0 target). 
In addition, although the TAM must vanish for a spin-0 target as a result of the spin sum rule, its spatial distribution may be non-trivial and, to the best of our knowledge, has so far not been investigated.

We find that the relativistic spatial distribution 
of transverse TAM in the transverse plane for a spin-0 target reads 
\begin{align}
&   \hspace{-0.5cm}
    J_{\perp,a}^{i}
    \left(\bm{b}_{\perp},P^z\right)
  = \int dr^{z}\;
    L_{\perp,a}^{i}
    \left(\bm{r},P^z\right) \cr
&   \hspace{-0.5cm}
  = \int \frac{d^{2}\Delta_\perp}{\left(2\pi\right)^{2}}\;
    e^{-i\bm{\Delta}_{\perp}\cdot\bm{b}_{\perp}}
    i \epsilon^{ij}_\perp
    \sqrt{\tau}
    X_{1}^{j}
    \left[
    4MP^{z}
    \frac{d}{dt}
    A_{a}
  + \frac{MP^{z}}{2\left(P^{0}\right)^{2}}
    D_{a}
    \right],
\label{Jk_spin0}
\end{align}
with the impact-parameter coordinate 
$\bm{b}_{\perp}=\left(\bm{r}_{\perp}-\bm{R}_{\perp},0\right)$ and $\epsilon^{12}_\perp=-\epsilon^{21}_\perp=1$.
We also introduced the rank-$n$ irreducible multipole tensors 
in the 2D momentum transfer space as 
$X_{0}=1$, $X_{1}^{i}=\Delta_{\perp}^{i}/\left|\bm{\Delta}_{\perp}\right|$,
$X_{2}^{ij}=\Delta_{\perp}^{i}\Delta_{\perp}^{j}/\left|\bm{\Delta}_{\perp}\right|^{2}-\delta^{ij}_\perp/2$~\cite{Kim:2022bia,Kim:2022wkc,Hong:2023tkv}.
As expected from the spin sum rule, we have
\begin{align}
    J^i_\perp(s',s)
 := \int d^{2}b_{\perp}\;
    J_{\perp}^{i}
    \left(\bm{b}_{\perp},P^z\right)
  = 0.
\end{align}
The result in Eq.~\eqref{Jk_spin0} shows that the transverse TAM in a spin-0 target is non-trivial at the distribution level, and that the transverse spin sum rule holds only after integration over the entire space.

\begin{figure*}[htbp]
  \centering
  \includegraphics[width=0.7\textwidth]{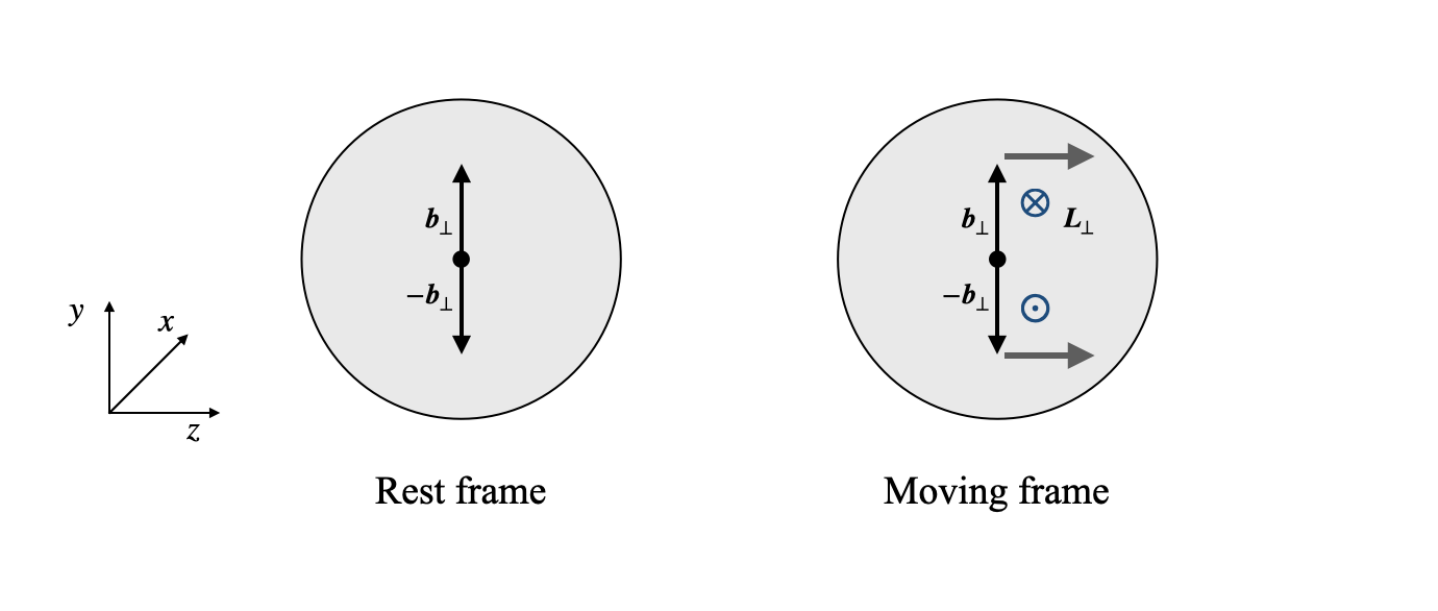}
  \hspace{0cm}
  \caption{
  Illustration of the transverse OAM associated with the first term in Eq.~\eqref{Jk_spin0} inside a spin-0 target. 
  The transverse vector $\bm{b}_{\perp}$ denotes a generic position in the transverse plane and the dark grey arrows represent the direction of the momentum. 
  When the target is boosted, each point inside the system acquires some momentum, thereby inducing some transverse OAM $\bm{L}_\perp$ contribution (represented by $\odot$ and $\otimes$ in the right panel). 
}

  \label{fig:1a}
\end{figure*}

\begin{figure*}[htbp]
  \centering
  \textbf{Transverse TAM distributions in a
pion}

  \vspace{0cm} 

  \includegraphics[width=0.9\textwidth]{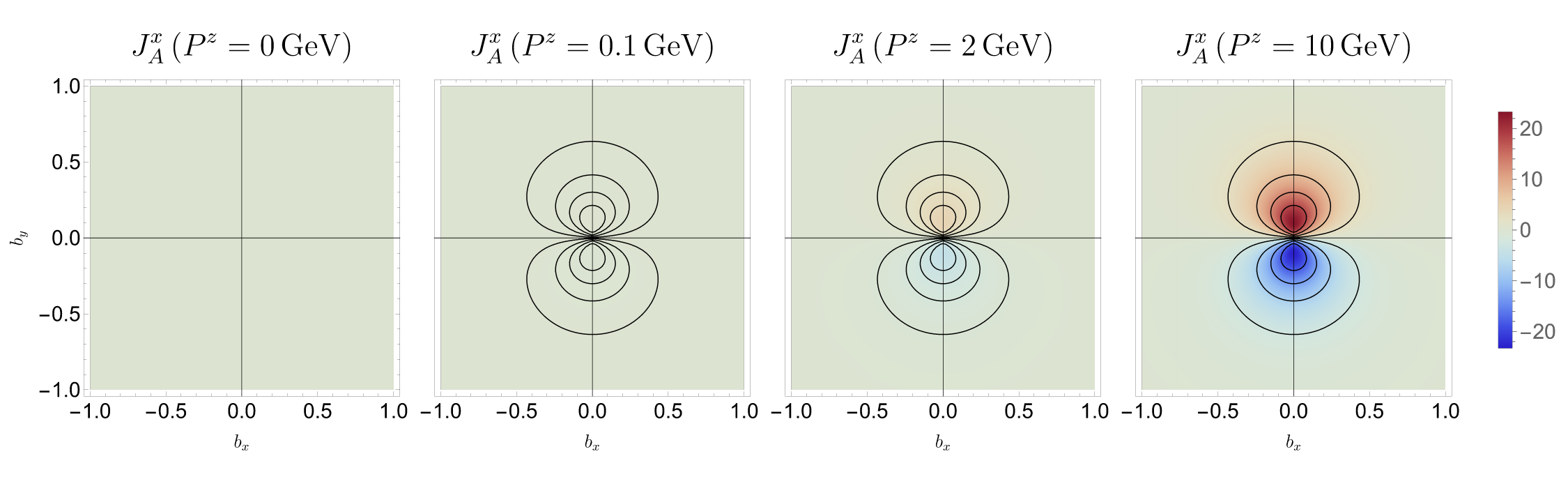}
  \hspace{0cm}

  \includegraphics[width=0.9\textwidth]{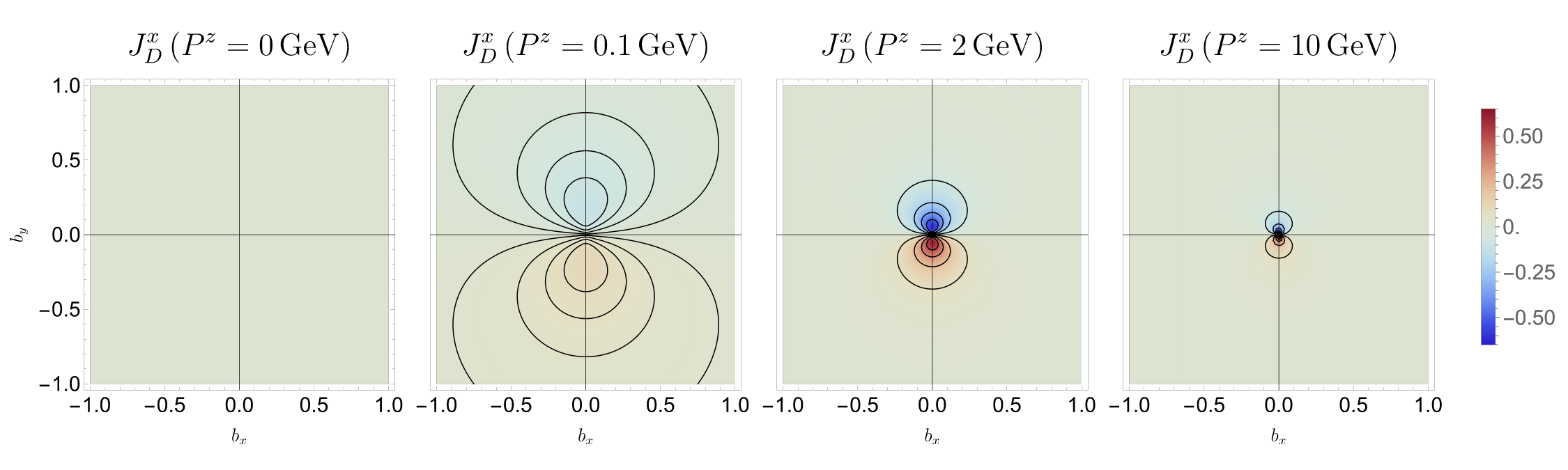}
  \hspace{0cm}

  \includegraphics[width=0.9\textwidth]{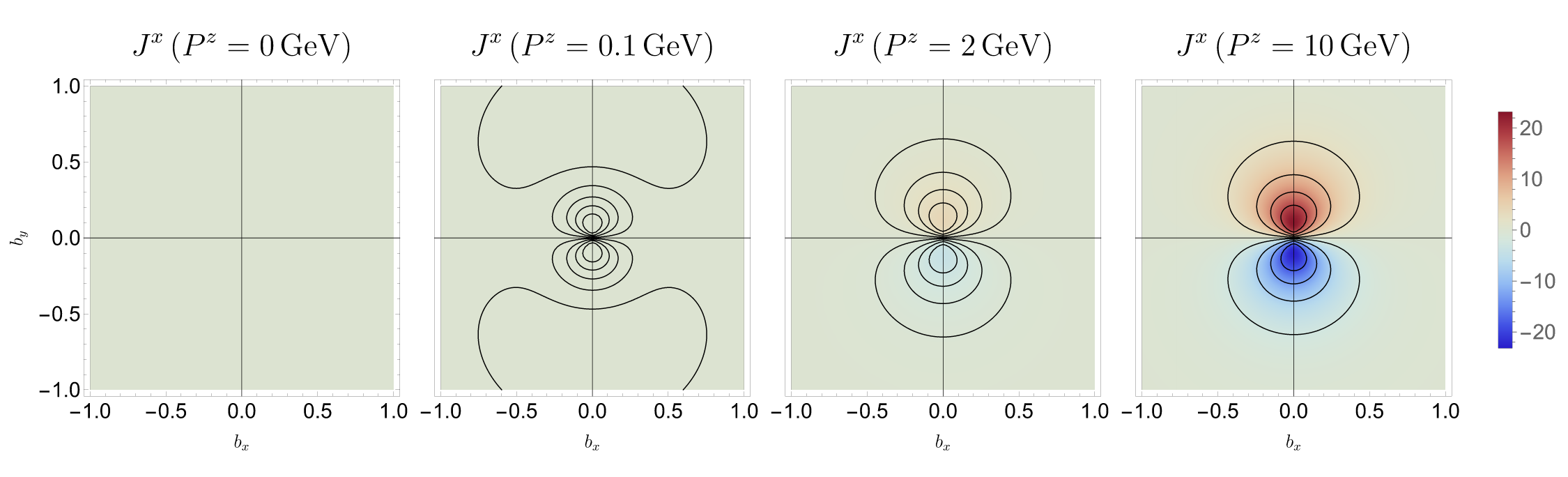}
  \hspace{0cm}

  \caption{Total (i.e.,~quark + gluon) relativistic spatial distributions of transverse TAM for a pion target (third line) and their individual contributions (first and second lines) according to Eq.~\eqref{Jk_spin0}, shown in the transverse plane for four different values of the pion momentum. 
  This illustration is based on the simple multipole model for the EMT form factors of Ref.~\cite{Hackett:2023nkr}, assuming a target mass of $M = 0.139$ GeV.
}

  \label{fig:2}
\end{figure*}

The first term in Eq.~\eqref{Jk_spin0} arises from the contribution $\epsilon^{ij3}r^jT^{03}$ and can be understood as follows. Let us imagine a classical, non-rotating sphere, see  Fig.~\ref{fig:1a}. When the sphere is moving with velocity $\bm{v}$, say in $z$-direction, each point of mass $dm=\rho(\bm{r})d^3r$ provides an OAM contribution $d\bm{L}=\bm{r}\times dm\,\bm{v}$ pointing in the transverse direction. Provided that we identify the origin of coordinates with the center of mass, the OAM integrated over the whole sphere will vanish by definition of the center of mass $\int d^3r\,\bm{r} \rho(\bm r)=\bm{0}$. The second term in Eq.~\eqref{Jk_spin0} comes from the contribution $\epsilon^{i3j}r^zT^{0j}$ that survives under integration over $r^z$ thanks to $\Delta^0=P^z\Delta^z/P^0$. This term is suppressed by two powers of $P^0$ with respect to the first one, which suggests that this is an induced relativistic correction. Since our current knowledge of hadron GFFs indicates that $D<0$ while $dA/dt>0$, the second term tends to partially counter the first term, as illustrated in the first two rows of Fig.~\ref{fig:2}. However, this correction is relatively small and vanishes in the IMF.

\subsection{Spin-1/2 Target}
The existence of intrinsic spin leads to
rather complicated dynamics inside the system. 
We find that the relativistic spatial distributions 
of transverse OAM and intrinsic spin in the transverse plane for a spin-1/2 target takes the form 
\begin{align}
&   L_{\perp,a}^{i} 
    \left(\bm{b}_{\perp},P^{z};s^{\prime},s\right)
  = \int dr^{z}\;
    L_{\perp,a}^{i} 
    \left(\bm{r},P^{z};s^{\prime},s\right)\cr
& = \int \frac{d^{2}\Delta_{\perp}}
    {\left(2\pi\right)^{2}}\,
    e^{-i\bm{\Delta}_{\perp}\cdot\bm{b}_{\perp}}
    \Bigg[
    \delta_{s^{\prime}s}\,
    i   \epsilon^{ij}_\perp   
    \sqrt{\tau}
    X_{1}^{j}  
    \tilde{L}_{1,a}^{U} \left(t\right)\cr
& + \left(\sigma_{\perp}\right)_{s^{\prime}s}^{i}
    X_{0}
    \tilde{L}_{0,a}^{T} \left(t\right)
  + \left(\sigma_{\perp}\right)_{s^{\prime}s}^{j}
    \tau 
    X_{2}^{ij} 
    \tilde{L}_{2,a}^{T}    \left(t\right)
    \Bigg]_{t=-\bm{\Delta}_{\perp}^{2}}
    \label{Lk4}
\end{align}
with a similar expression for $S^i_{\perp,q}$.
The amplitudes $\tilde{L}_{n}$ and $\tilde{S}_{n}$ are the rank-$n$ multipole amplitudes contributing, respectively, to the transverse OAM and intrinsic spin distributions. 
Here, the superscripts $U$ and $T$ denote unpolarized and transversely polarized contributions, respectively. We note that there is no contribution proportional to $\sigma^z_{s's}$, meaning that the distributions for unpolarized and longitudinally polarized spin-$\frac{1}{2}$ targets are the same.

The explicit expressions for the multipole amplitudes are
\begin{align}
    \tilde{L}_{1,a}^{U}
& = \frac{d}{dt}    
    \left[
    \frac{4MP^{z}\left[P^{0}+M\left(1+\tau\right)\right]}
    {P^{0}+M}
    A_{a}   
    \right] \cr
& + \frac{MP^{z}\left[P^{0}+M\left(1+\tau\right)\right]}
    {2\left(P^{0}\right)^{2}\left(P^{0}+M\right)}
    D_{a}    
  - \frac{MP^{z}}{2P^{0}\left(P^{0}+M\right)}
    L_{a} \cr
& + \frac{d}{dt}  
    \left[
    tP^{z}
    \left(
    \frac{L_{a}}{P^{0}+M}
  + \frac{
    J_{a}   
  + S_{a}   
    }{P^{0}}
    \right)
    \right],   
    \label{LU1}\\
    \tilde{L}_{0,a}^{T}
& = - \frac{d}{dt}    
    \left[
    \frac{\left(P^{z}\right)^{2}}{2M\left(P^{0}+M\right)}
    t
    A_{a}   
    \right] 
  - \frac{\left(P^{z}\right)^{2}}
    {16M\left(P^{0}\right)^{2}\left(P^{0}+M\right)}
    t
    D_{a} \cr
& 
  + \frac{M\left[P^{0}+M\left(1+\tau\right)\right]}
    {2P^{0}\left(P^{0}+M\right)}
    L_{a}
  + \frac{1}{2}
    \frac{d}{dt}  
    \left[
    t
    L_{a}
    \right]\cr
&+ \frac{d}{dt}  
    \left[
    \frac{t\left(P^{z}\right)^{2}}{2M}
    \left(
    \frac{L_{a}}{P^{0}+M}
  + \frac{
    J_{a}  
  + S_{a}   
    }{P^{0}}
    \right)
    \right],
    \\
    \tilde{L}_{2,a}^{T}
& = - \frac{d}{dt}    
    \left[
    \frac{4M\left(P^{z}\right)^{2}}{P^{0}+M}
    A_{a}   
    \right]
  - \frac{M\left(P^{z}\right)^{2}}
    {2\left(P^{0}\right)^{2}\left(P^{0}+M\right)}
    D_{a}  
  + 4M^{2}
    \frac{d}{dt}
    L_{a}\cr
& + \frac{d}{dt}
    \left[
    4M\left(P^{z}\right)^{2}
    \left(
    \frac{L_{a}}{P^{0}+M}
  + \frac{J_{a}+S_{a}}{P^{0}}
    \right)
    \right],    \\
    \tilde{S}_{1,q}^{U}
& = - \frac{MP^{z}}{2P^{0}\left(P^{0}+M\right)}
    \, G_A^q,  \\
    \tilde{S}_{0,q}^{T}
& = \frac{\left(P^{0}+M\right)^{2}-\left(P^{z}\right)^{2}}
    {4P^{0}\left(P^{0}+M\right)}
    \, G_A^q 
  + \frac{t}{16MP^{0}}
    G_{P}^{q},   \\
    \tilde{S}_{2,q}^{T}
& = - \frac{M^{2}}{2P^{0}\left(P^{0}+M\right)}
    \, G_A^q
  - \frac{M}{2P^{0}}
    G_{P}^{q}
    \label{ST2},
\end{align}
where we introduced for convenience the OAM form factor $L_{a}=J_{a}-S_{a}$. 
One might expect that setting the AM form factors to zero, $J_a,L_a, S_a \to 0$, and turning off the nucleon spin matrices, $\bm{\sigma} \to \bm{0}$, would yield the spin-0 TAM distribution in Eq.~\eqref{Jk_spin0}. However, as observed in Ref.~\cite{Lorce:2020onh,Chen:2022smg} for the electromagnetic current, this is not sufficient since one has also to (i) account for the fact that the fermionic bilinear $\bar u' u$ corresponds to a factor $2\sqrt{P^2}$ in the bosonic amplitude, which amounts to $(A,D)|_\text{fermion}\to \frac{1}{\sqrt{1+\tau}} (A,D)|_\text{boson}$, and  (ii) switch off Wigner rotation effects~\cite{Wigner:1932eb,Melosh:1974cu}, i.e.,~take the limits $\cos\theta=\frac{P^0+M(1+\tau)}{(P^0+M)\sqrt{1+\tau}}\to 1$ and $\sin\theta=-\frac{\sqrt{\tau}P^{z}}{(P^0+M)\sqrt{1+\tau}}\to 0$.

The transverse OAM and intrinsic spin distributions in Eq.~\eqref{Lk4} coincide exactly in the BF limit, i.e.,~$P^{z}=0$, with the projections of the BF distributions presented in Ref.~\cite{Lorce:2017wkb} onto the transverse plane.
Additionally, we reproduce the unpolarized transverse intrinsic spin distribution for arbitrary value of $P^z$ obtained in Ref.~\cite{Lorce:2018egm}, and generalize the result to polarized nucleon targets.

In this work, 
we denote the distributions for an unpolarized target as~\cite{Lorce:2018egm,Won:2025dgc}
\begin{align}
    g_{a}
    \left(\bm{b}_{\perp},P^{z}\right)
&:= \frac{1}{2} 
    \sum_{s^{\prime},s}
    g_{a}
    (\bm{b}_{\perp},P^{z};s^{\prime},s)
    \delta_{s^{\prime}s}, 
\end{align}
with $g=L^{i},S^{i},J^{i}$. 
The distributions for a transversely
polarized target along the $x$-axis are given as 
\begin{align}
    g_{a}^{T}
    \left(\bm{b}_{\perp},P^{z}\right)
&:= g_{a}
    \left(\bm{b}_{\perp},P^{z};s_{x},s_{x}\right)  
\end{align}
with the transversely polarized spin-1/2 states 
$\ket{s_{x}=\pm\frac{1}{2}}=\left(\ket{s^{z}=+\frac{1}{2}}\pm\ket{s^{z}=-\frac{1}{2}}\right)/\sqrt{2}$.
We remind that the transverse OAM and intrinsic spin distributions for a longitudinally polarized target are identical to those for an unpolarized target.
Thus, we do not discuss them separately.
Moreover, analogously to Fig.~\ref{fig:2} showing the transverse TAM distributions in a spin-0 target, we restrict our discussion of the transverse AM distributions in a spin-1/2 target to the range $P^{z}\leq 10$ GeV, as those distributions diverge when $P^{z}\to\infty$.

\begin{figure*}[htbp]
  \centering
  \textbf{Transverse OAM, intrinsic spin, and TAM distributions in an unpolarized nucleon}

  \vspace{0cm} 

  \includegraphics[width=0.9\textwidth]{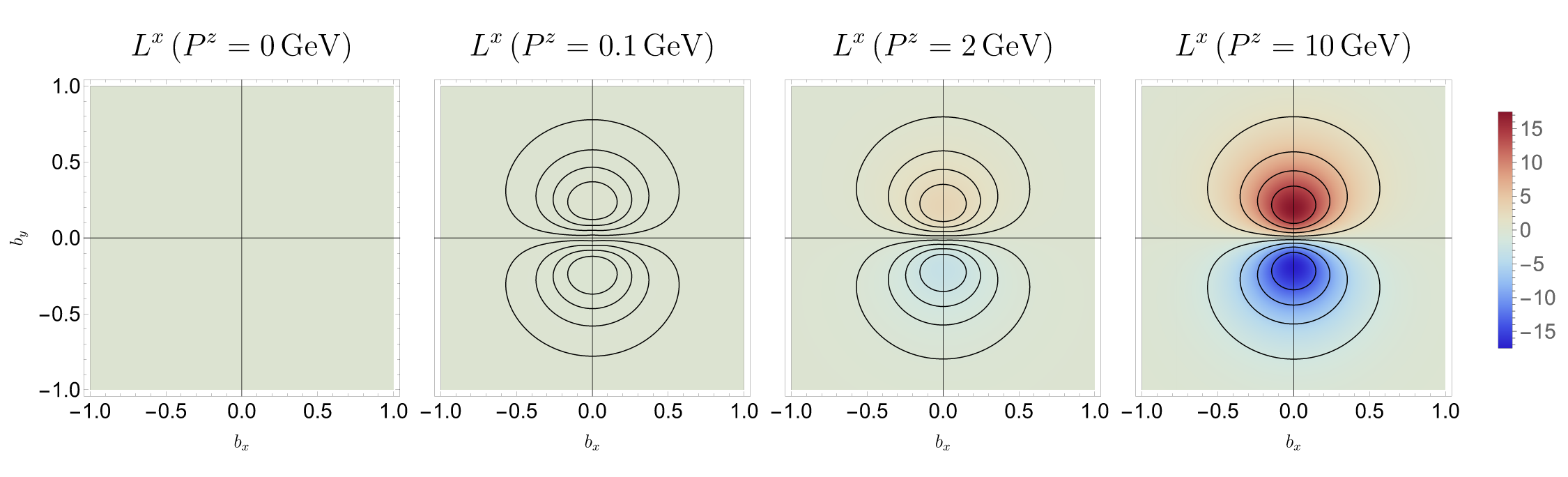}
  \hspace{0cm}

  \includegraphics[width=0.9\textwidth]{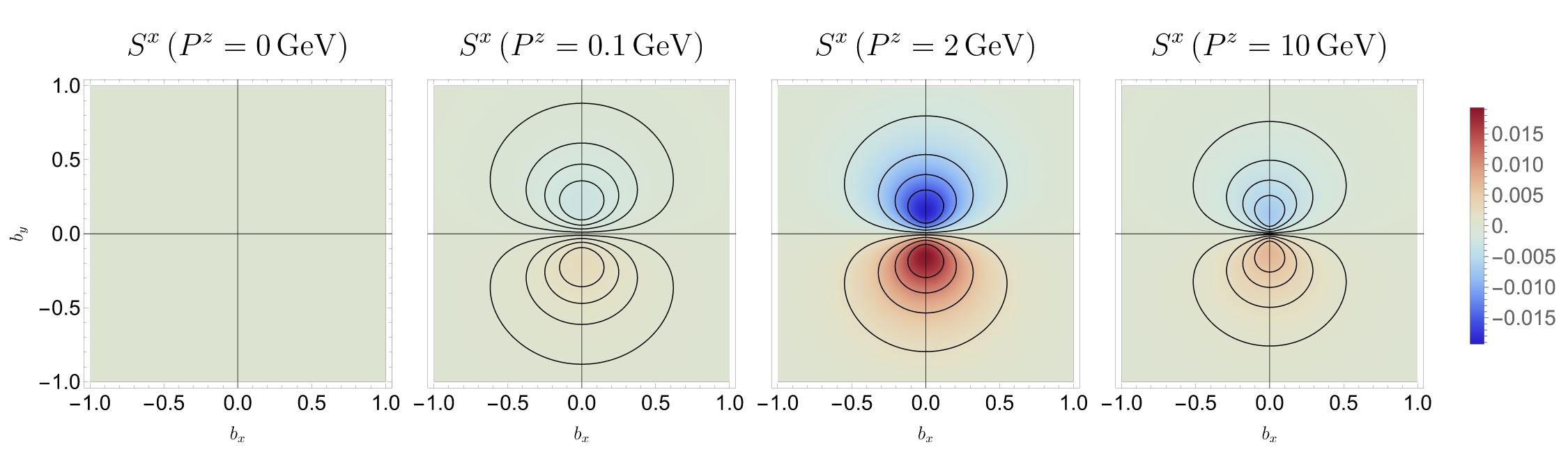}
  \hspace{0cm}

  \includegraphics[width=0.9\textwidth]{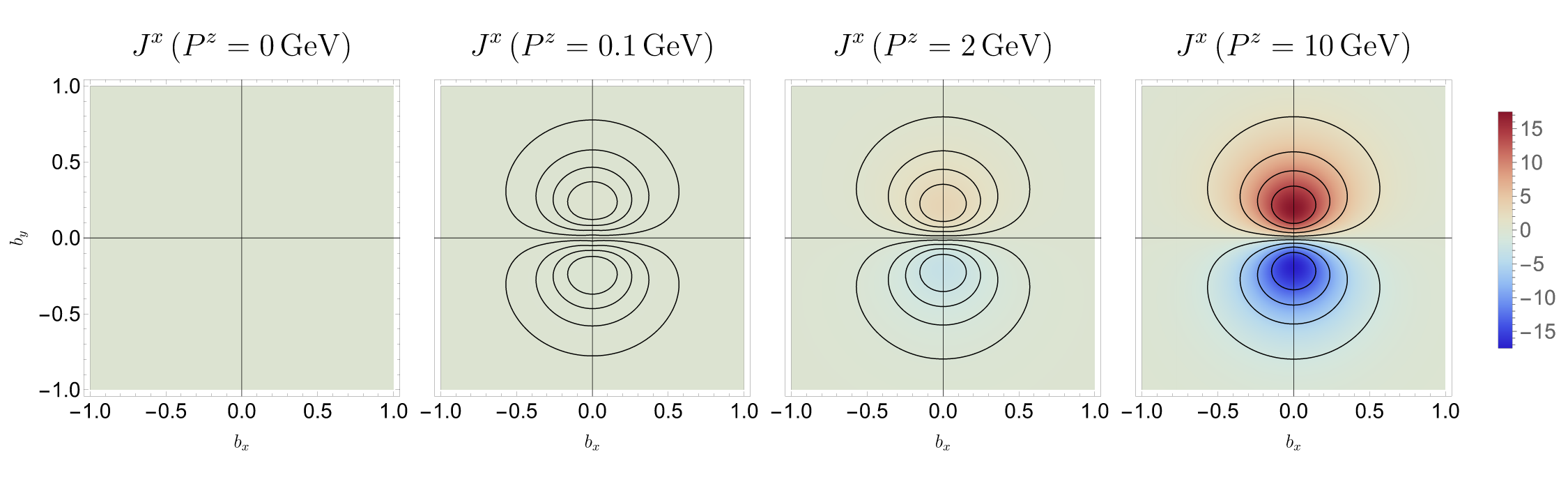}
  \hspace{0cm}

  \caption{Total (i.e.,~quark + gluon) relativistic spatial distributions of transverse OAM, spin, and TAM in the transverse plane 
  for an unpolarized nucleon for four different values of the nucleon momentum. This illustration is based on the simple multipole model for the EMT form factors of Ref.~\cite{Hackett:2023rif}, assuming a target mass of $M = 0.938$ GeV.
  }
  \label{fig:3}
\end{figure*}

\begin{figure*}[htbp]
  \centering
  \textbf{Transverse OAM, intrinsic spin, and TAM distributions in a transversely polarized nucleon}

  \vspace{0cm} 

  \includegraphics[width=0.9\textwidth]{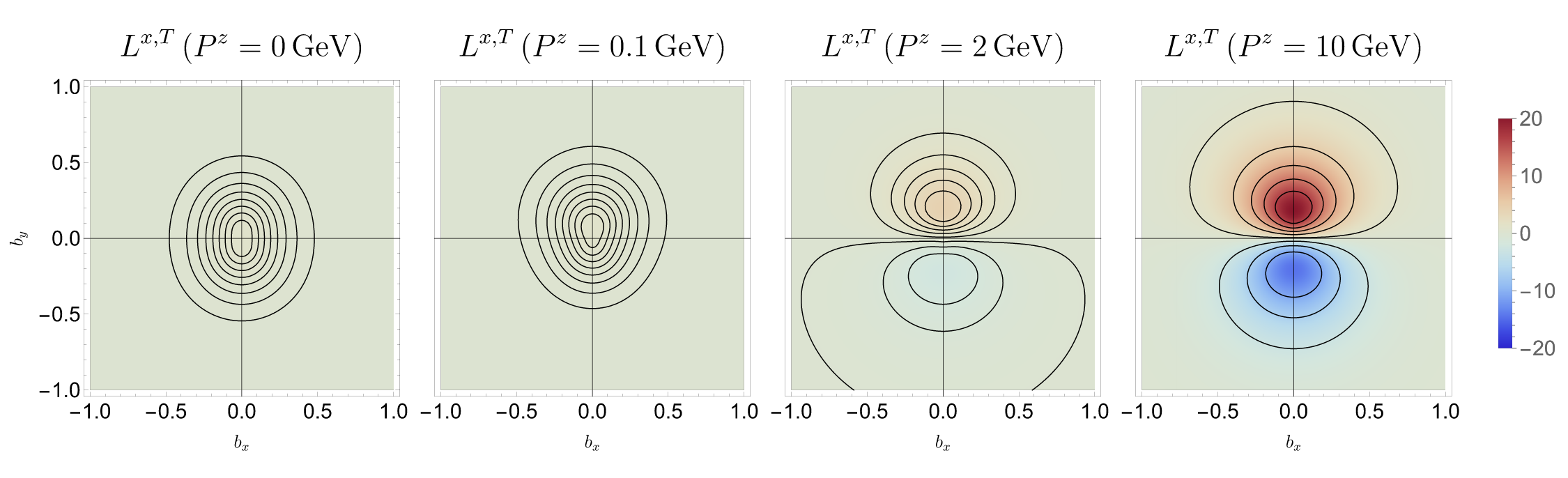}
  \hspace{0cm}

  \includegraphics[width=0.9\textwidth]{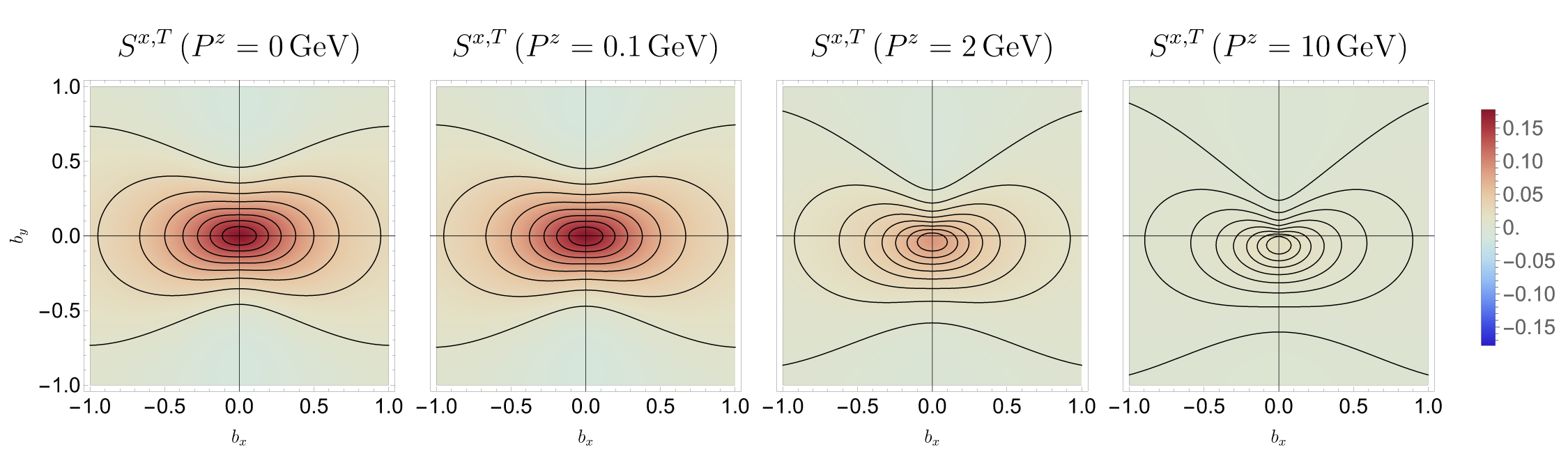}
  \hspace{0cm}

  \includegraphics
  [width=0.9\textwidth]{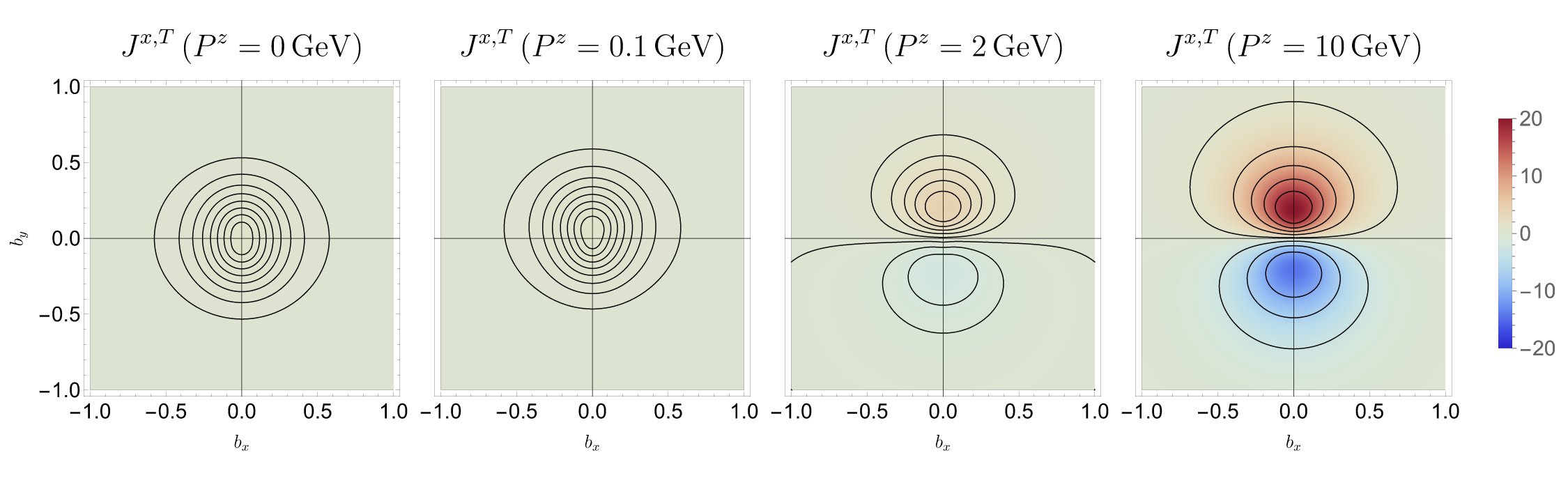}
  \hspace{0cm}

  \caption{Total (i.e.,~quark + gluon) relativistic spatial distributions of transverse OAM, spin, and TAM in the transverse plane 
  for a transversely polarized nucleon along the $x$-axis for four different values of the nucleon momentum. This illustration is based on the simple multipole model for the EMT form factors of Ref.~\cite{Hackett:2023rif}, assuming a target mass of $M = 0.938$ GeV.}

  \label{fig:4}
\end{figure*}

In Fig.~\ref{fig:3},
we show the total (i.e.,~quark + gluon) contributions
to the relativistic spatial distributions 
of transverse OAM, intrinsic spin, and TAM in the transverse plane 
for an unpolarized nucleon at various values of $P^{z}$. 
We observe that dipolar structures are induced by the motion of the target.
Similar to the spin-0 case,
the unpolarized transverse AM distributions are antisymmetric under reflection about the $x$-axis, and hence vanish once integrated over all space. 
In Fig.~\ref{fig:4},
we show the same transverse distributions, now for a transversely polarized nucleon along the $x$-axis. On top of the dipole contribution associated with the spin-independent part, there are monopole and quadrupole contributions associated with the spin-dependent part, see Eq.~\eqref{Lk4}.

The transverse OAM and intrinsic spin distributions are normalized as 
\begin{align}
&   L^i_\perp(P^z;s',s)
 := \int d^{2}b_{\perp}\;
    L_{\perp}^{i} 
    \left(\bm{b}_{\perp},P^{z};s^{\prime},s\right)
    \cr
& = \left(\sigma_{\perp}\right)_{s^{\prime}s}^{i}
    \left[
  - \frac{\left(P^{z}\right)^{2}}{2M\left(E_{P}+M\right)}
    A  (0)
  + \frac{E_{P}}{M}
    J  (0)
  - \frac{M}{E_{P}}
    S  (0)
    \right], 
    \label{Lk5}\\
&   S^i_\perp(P^z;s',s)
 := \int d^{2}b_{\perp}\;
    S_{\perp}^{i}
    \left(\bm{b}_{\perp},P^{z};s^{\prime},s\right)\cr
& = \left(\sigma_{\perp}\right)_{s^{\prime}s}^{i}
    \frac{M}{E_{P}}
    \,\frac{G^q_A(0)}{2},
    \label{Sk5}
\end{align}
with $E_{P}=P^{0}|_{t=0}=\sqrt{\left(P^{z}\right)^{2}+M^{2}}$.
Combining Eqs.~\eqref{Lk5} and \eqref{Sk5}, 
we obtain the transverse spin sum rule for a spin-1/2 target 
\begin{align}
    J^i_\perp(P^z;s',s)
& = L^i_\perp(P^z;s',s)
  + S^i_\perp(P^z;s',s)
  = \frac{\left(\sigma_{\perp}\right)_{s^{\prime}s}^{i}}{2},
\label{CanSpinSumRule}
\end{align}
where we used the relations~\eqref{spinrelation} and the EMT form factor constraints~\eqref{GFFconstraints} . 
We reproduce in particular the transverse spin sum rule about the relativistic center of spin discussed in Refs.~\cite{Leader:2011cr,Lorce:2021gxs} and confirm its $P^z$-independence. 
However, our results are more general since they provide a decomposition of the transverse TAM into orbital and intrinsic spin contributions. 
Note that the latter split is $P^z$-dependent, with the transverse TAM becoming more and more orbital-like as we increase the momentum.
Note that the $P^z$-independence of TAM is not true if it is defined relative to the center of mass or energy of the system~\cite{Ji:2020hii,Lorce:2021gxs,Lorce:2018zpf}. 
In an upcoming work~\cite{Lorce:2025ll}, 
we intend to study the effect of different pivot choices on the spatial distributions of transverse AM.

\section{Conclusions \label{sec.4}}
In this work, we reviewed the generalized AM tensor and EMT operators in QCD, and defined in general the corresponding AM distributions.
We briefly explained the technical difficulties that are encountered in defining relativistic spatial distributions of transverse OAM in the transverse plane.
Adopting the quantum phase-space formalism, 
we derived relativistic spatial distributions in a generic Lorentz frame characterized by non-zero target momentum in the 3D space.
Projecting these 3D distributions onto 2D ones, i.e., by integrating over the longitudinal direction, we obtained the relativistic spatial distributions of transverse OAM and intrinsic spin in the transverse plane.

We then analyzed how the relativistic spatial distributions of transverse OAM, intrinsic spin, and TAM evolve as spin-0 and spin-1/2 targets are boosted along the longitudinal axis. 
For spin-0 targets, the spin sum rule is trivially satisfied, but the spatial distribution reveals a non-trivial structure that arises purely from Lorentz boost effects. 
Since this is a spin-independent feature, it also appears in higher-spin targets and plays a key role in understanding how the spatial distribution of angular momentum changes across different frames.
Thus, we also discussed the origin and interpretation of each contribution to the transverse OAM distribution in a spin-0 target.
For spin-1/2 targets, relativistic intrinsic spin dynamics complicates the internal structure.
We examined the transverse OAM, intrinsic spin, and TAM distributions for both unpolarized and transversely polarized spin-1/2 targets.
Interestingly, for longitudinally polarized spin-1/2 targets, the transverse AM distributions exactly match those of unpolarized targets.
Finally, we verified that the transverse spin sum rule is satisfied by integrating over the spatial coordinates the transverse TAM distributions in a transversely polarized spin-1/2 target.

\section*{Acknowledgments}
H.-Y.W. acknowledges the French government (Ministère de l’Europe et des Aﬀaires Étrangères) for the France Excellence scholarship through Campus France funded, Grant No. 141295X. 
All authors also thank the SCPP project RD/0523-IOE00I0-086 from IRCC, IIT Bombay for funding. 

\bibliography{PLB_TOAM}

\end{document}